\documentclass[twoside,10pt]{article}
\usepackage{1}

\title{\Large \bf Overview of ALICE results}
\author{E. L. Kryshen for the ALICE Collaboration}
\date{}

\def\jpsi{J$/\psi$ }

\begin{document}
\maketitle

\begin{center}
\vspace*{-0.3cm}
{\it CERN, CH-1211 Geneva 23, Switzerland}
\end{center}

\vspace{0.3cm}

\begin{center}
{\bf Abstract}\\
\medskip
\parbox[t]{10cm}{\footnotesize
Selected ALICE results on the global event properties, particle spectra, azimuthal anisotropy, heavy flavour and quarkonium production in Pb--Pb collisions at $\sqrt{s_{NN}} = 2.76$ TeV are presented. First results on p--Pb collisions at $\sqrt{s_{NN}} = 5.02$ TeV are briefly reviewed.}
\end{center}

ALICE (A Large Ion Collider Experiment) is aimed to study hot and dense QCD matter produced in heavy ion collisions at LHC~\cite{alice}. ALICE collected about 10 $\mu$b$^{-1}$ and 100 $\mu$b$^{-1}$ of Pb--Pb collisions at $\sqrt{s_{NN}} = 2.76$ TeV during the first heavy ion runs in 2010 and 2011, respectively. In the beginning of 2013, LHC delivered 30 nb$^{-1}$ of p--Pb collisions at $\sqrt{s_{NN}} = 5.02$ TeV, important reference data for the Pb--Pb studies. Selected results based on these data samples are briefly summarised in the following.

ALICE measurements of the global event observables indicate that the matter, produced in Pb--Pb collisions at LHC, reveal even more extreme properties than at lower energies. The charged-particle density at mid rapidity amounts to $dN/d\eta \approx 1600$ in central Pb--Pb collisions at LHC, factor 2.2  higher than in central Au-Au collisions at RHIC~\cite{multiplicity}. It corresponds to an initial energy density of about 15 GeV/fm$^{3}$ (at a conventional value of 1 fm/$c$ for the thermalization time), factor 3 higher than in Au--Au collisions at the top energy of RHIC. The measured slope of the direct photon spectrum, $T = 304\pm 51$ MeV~\cite{temperature}, suggests that the initial temperature of the produced medium goes well above the critical temperature of 150--160 MeV predicted for the deconfinement state transition by lattice QCD calculations. The volume of the produced fireball, measured with two-pion Bose-Einstein correlations, increases by factor of two from RHIC to LHC, while the matter
lifetime, roughly proportional to the longitudinal dimension of the fireball, turns out to be more than 10 fm/$c$, 20\% higher than at RHIC, in line with hydrodynamic predictions~\cite{hbt}.

Further constraints on the evolution of the produced medium come from the measurement of the momentum-space anisotropy (flow) which is quantified by the Fourier decomposition of azimuthal particle distributions. Large elliptic flow (or second Fourier coefficient, $v_2$), observed both at RHIC and LHC energies~\cite{flow}, appears to be consistent with the dynamics of an almost perfect liquid characterized by the shear viscosity to entropy density ratio $\eta/s$ close to the lower bound of $1/4\pi$ from AdS/CFT. Furthermore, the elliptic flow of identified particles reveals a clear mass ordering at low $p_{\rm T}$~\cite{noferini}, being explained by the strong collective dynamics of the medium. However, the Number of Constituent Quark (NCQ) scaling of the elliptic flow, observed at RHIC and considered as a direct consequence of the coalescence hadronization mechanism, is not that good at LHC energies~\cite{noferini}. Significant triangular flow (third Fourier coefficient, $v_3$), measured by ALICE~\cite{
flowpid}, appears to be highly sensitive to the $\eta/s$ ratio and initial state fluctuations providing promising tools to constrain hydrodynamic models.

ALICE measurements of the low-momentum proton, pion and kaon $p_{\rm T}$ spectra in central Pb--Pb collisions agree with hydrodynamic predictions within 20\% supporting hydrodynamic interpretation of the data at LHC~\cite{spectra}. Hydro-inspired blast wave fits to these spectra allowed to extract mean collective velocity of the transverse expansion which was found to be about 65\% of speed of light, 10\% higher than at RHIC and in good agreement with the observed tendency from the RHIC energy scan.

ALICE also measured integrated particle yields for various particle species. At lower energies, the yields were surprisingly well described in terms of a simple thermal model with a common chemical freeze-out temperature $T_{\rm ch}$~\cite{andronic}.  However, thermal fits for  0--20\% central collisions at LHC provide unexpectedly low $T_{\rm ch} = 152 \pm 3$ MeV and fail to describe the yields of multistrange hyperons~\cite{bellini}. On the other hand, a model with $T_{\rm ch}= 164$ MeV, extrapolated from the RHIC data, instead seems to agree with the ratios involving multistrange hyperons, but missing $p/\pi$ and $\Lambda/\pi$ ratios. Arguably, the significant deviation from the thermal model can be explained by the final-state interactions in the hadronic phase.

High-$p_{\rm T}$ hadrons are produced in hard interactions at early stages of heavy ion collisions and can be used as effective tomography probes of the produced medium. Energy loss of  hard partons in the medium may result in a strong suppression of high-$p_{\rm T}$ hadrons (jet quenching) which was indeed observed at RHIC and quantified in terms of the nuclear modification factor $R_{AA}$ ($p_{\rm T}$ spectra in AA collisions normalized to appropriately scaled pp spectra). The suppression of high-$p_{\rm T}$ hadrons appeared to be even stronger in central Pb--Pb collisions at LHC with $R_{AA}$ reaching minimum of about 0.14 for $p_{\rm T} \approx 6$ GeV/$c$ and slowly increasing at high $p_{\rm T}$~\cite{raa}.

The suppression for open heavy flavour $D^0$, $D^+$ and $D^{*+}$ mesons reaches factor 5 at $p_{\rm T} \sim 10$ GeV/$c$~\cite{draa}, almost as large as that observed for light hadrons (dominated by pions from gluon fragmentation) providing an indication of no strong colour charge or mass dependence of the in-medium energy loss. The observed elliptic flow of prompt $D^{0}$ mesons is also comparable with $v_2$ of light hadrons~\cite{dflow} suggesting that the azimuthal anisotropy of the system is effectively transferred to charm quarks via multiple interactions in the medium.

Suppression of hidden charm mesons due to colour-screening effects was one of the first signals predicted for the formation of deconfined phase and indeed observed at SPS and RHIC. However, high abundance of charm quarks at LHC may also result in an enhancement of bound $c\bar{c}$ states via regeneration in thermalized QGP medium.  The \jpsi suppression, measured by ALICE versus number of participants $N_{\rm part}$, appeared to flatten at $\langle  N_{\rm part}\rangle \sim 100$ being much weaker than at RHIC for central collisions~\cite{jpsiRaaForward}. Such a centrality dependence and additional rapidity and $p_{\rm T}$ differential studies suggest that $c\bar{c}$ regeneration processes indeed play an important role at LHC energies. The observed hint for a non-zero elliptic flow for J/$\psi$ in semi-central Pb-Pb collisions is also in favour of this picture~\cite{jpsiflow}.

\jpsi production has been also measured at forward and mid rapidity in ultraperipheral collisions (UPC) which are dominated by photon-induced reactions~\cite{coherent-forward,coherent-central}. In the LO pQCD, coherent \jpsi photoproduction cross section is proportional to the squared nuclear gluon density providing a direct tool to study poorly known gluon shadowing in nuclei at small $x \sim 10^{-3}-10^{-2}$, one of the most important initial state effects in heavy ion collisions.  ALICE measurements appear to be in good agreement with a model which incorporates nuclear gluon shadowing according to EPS09LO global fits~\cite{Eps09}.

Proton-nucleus collisions provide futher tools to study initial and final state effects in cold nuclear matter and establish a baseline for the interpretation of heavy-ion results. The pseudorapidity dependence of the charged particle density in non-single-diffractive p--Pb events, measured by ALICE~\cite{pAmult}, is well described by DPMJET and the HIJING 2.1 generator with gluon shadowing tuned to describe RHIC d--Au data and consistent with EPS09 fits. Gluon saturation models predicted a steeper pseudorapidity dependence. The nuclear modification factor $R_{\rm pPb}$ of charged particles is consistent with unity at transverse momentum above 2 GeV/$c$ indicating that the strong suppression of hadron production measured in Pb--Pb collisions at LHC is not an initial state effect but is a consequence of jet quenching in hot QCD matter~\cite{pArpA}. ALICE also measured the \jpsi suppression pattern in p--Pb collisions, an important baseline for the interpretation of the \jpsi suppression in Pb--Pb~\cite{
pAjpsi}. The results are in agreement with models incorporating EPS09 shadowing or coherent parton energy losses, while Color-Glass condensate predictions are disfavoured by this measurement.

p--Pb collisions also appeared to be good for surprises. The analysis of two-particle angular correlations in high-multiplicity p--Pb collisions showed the presence of a ridge structure elongated in the pseudorapidity direction, so-called near-side ridge. Subtraction of correlation pattern for low-multiplicity events revealed a symmetric  structure on the away side, similar to modulations caused by elliptic flow in Pb--Pb~\cite{pAcorr}. The dependence of $v_2$ coefficient, corresponding to these modulations, on $p_{\rm T}$ for identified particles exhibit a mass ordering pattern similar to Pb--Pb in agreement with hydrodynamic models~\cite{pAflow}. Other models attribute the effect to gluon saturation in Pb or to parton-induced final-state effects.

In conclusion, the ALICE collaboration obtained a wealth of interesting physics results from the first heavy ion runs at LHC revealing many new phenomena not observed at lower energies. ALICE is entering a charm era of precision measurements and is looking forward to new discoveries in Pb--Pb collisions at higher energy.

\end{document}